\documentclass[aps,twocolumn]{revtex4-1}
\usepackage{graphicx,epsfig}
\usepackage{amsmath}
\usepackage[T1]{fontenc}
\usepackage{hyperref}
\hypersetup{
     colorlinks   = true,
     citecolor    = magenta,
     linkcolor    = blue,
}

\begin{document}

\title{Spin-1 Bose Hubbard model with nearest neighbour extended
  interaction} \author{Sk Noor Nabi}%\email{sk.noor@iitg.ernet.in}
\author{Saurabh Basu} \email{saurabh@iitg.ernet.in}
\affiliation{Department of Physics, Indian Institute of Technology
  Guwahati, Guwahati, Assam 781039, India} \date{\today}

\begin{abstract}
We have studied a spinor ($F=1$) Bose gas in presence of the density-density interaction through the mean field approach and the perturbation theory for either sign of the spin dependent interaction, namely the antiferromagnetic (AF) and the ferromagnetic cases. In the AF case, the
charge density wave (CDW) phase appears to be sandwiched between the Mott insulating
(MI) and the supersolid phases for small values of the
extended interaction strength. But the CDW phase completely occupies
the MI lobe when the extended interaction strength is larger than a
certain critical value related to the width of the MI lobes and hence opens up the possibilities of spin singlet and nematic CDW insulating phases.  In the ferromagnetic
case, the phase diagram shows similar features as that of the AF
case and are in complete agreement with a spin-0 Bose gas. The perturbation expansion calculations nicely corroborate the mean field phase results in both these cases. Further, we extend our calculations
in presence of a harmonic confinement and obtained the momentum
distribution profile that is related to the absorption spectra in order to distinguish between different phases.
\end{abstract}


\maketitle

\section{Introduction}
\label{intro}
The cooling and trapping of neutral alkali atoms in optical lattices,
formed by two or more counter propagating laser beams make
it possible to explore a plethora of quantum many body phenomena compared to its
condensed matter counterpart due to adequate control over various
experimental parameters. The general features of the cold atoms trapped in
optical lattices were first theoretically described by the Bose Hubbard
model (BHM) \cite{PhysRevLett.81.3108}, where the superfluid (SF) to
Mott insulator (MI) transition can be achieved by tuning a competition between
the tunneling to the interatomic interaction potential. As a result of rapid technological
improvement, the first experimental signature of the SF-MI transition
was observed by Greiner $et$ $al.$ for a magnetically
trapped scalar Bose gas in an optical lattice \cite{Greiner}.  \\ \indent Despite
success in magnetic trapping of a scalar or spin-0 Bose gas, efforts to study the
spinor Bose gas have gained much more momentum after the MIT group
successfully confined $^{23}Na$ spin-1 condensate by using an optical
dipole trap \cite{PhysRevLett.80.2027}. The optical trap which
interacts via the electric fields of the laser beams with the dipole force
of the neutral atoms helps in distinguishing all the hyperfine spin degrees of
freedom and thus they are called as a gas of spinor bosons. Since then, several theoretical
and experimental attention have been paid to study the spin-1 \cite{PhysRevLett.81.5257,PhysRevLett.82.2228,PhysRevLett.84.4031,PhysRevLett.87.080401,PhysRevLett.100.180403} as well
as spin-2 \cite{PhysRevLett.92.040402,PhysRevLett.92.140403,PhysRevA.67.063607} Bose gases which have rich ground state structures consisting
of antiferromagnetic (or polar) and ferromagnetic for the former one \cite{Kawaguchi,RevModPhys.85.1191}, while
another additional cyclic phase \cite{Kawaguchi,RevModPhys.85.1191,PhysRevA.65.063602} appears for the latter.  \\ \indent In
this work, we primarily focus on the spin-1 ($F=1$) Bose gas whose
general properties were first theoretically analyzed by Ho
\cite{PhysRevLett.81.742} and Machida \cite{ohmi} to illustrate the spin
textures and topological excitations where
the spinor components transform to each other in the spin space via a
spin-gauge rotational symmetry. Besides, different types of the MI
phases, including the spin singlet, nematic as well as exotic
fractionalized phase that break both the spin and charge symmetry and
the SF phases were studied in Refs.\cite{PhysRevLett.88.163001,PhysRevA.68.063602}. Later,
possible ground state structures of the spin nematic and spin singlet
MI phases and the transition between them were investigated in
Refs.\cite{PhysRevB.69.094410,PhysRevA.68.063602}. Further the existence of
the dimerized phase is explored using an effective spin Hamiltonian in Refs.\cite{PhysRevLett.90.250402,EPL.63.505}.  \\ \indent Apart from all these
activities, a large number of review articles on the spinor Bose gas exist that emphasizes
the studies in presence of disorder \cite{PhysRevA.83.013605,NoorJPB}, external
magnetic field through the linear \cite{PhysRevA.86.063614,PhysRevLett.93.120405,PhysRevA.93.053628} and
quadratic Zeeman strengths \cite{PhysRevA.86.063614,PhysRevA.92.043617,PhysRevA.93.033633,PhysRevA.93.063607,PhysRevLett.107.195306}, spin-orbit
couplings (SOC) \cite{PhysRevA.86.043602,PhysRevA.91.023608,PhysRevA.93.013629,PhysRevA.93.023615,PhysRevLett.117.125301} and synthetic magnetic fields \cite{NoorEPL} etc. Among them, the inclusion
of SOC after its recent experimental realization using Raman coupling
between hyperfine levels \cite{arXiv:1501.05984} gives rise to more than one
minima in the single particle dispersion relation which leads to
different exotic ground state structures like plane and standing wave
\cite{PhysRevA.86.043602} and various striped ferromagnetic phases
\cite{PhysRevA.91.023608}. Also usages of the hyperfine spin states as
short lattice dimension, known as the synthetic dimension \cite{PhysRevLett.112.043001}, to create
spatially varying SOC gives rise to multiple density ordered SF phases
such as the charge density or the spin density wave phases
\cite{PhysRevA.94.063613}.  \\ \indent Although the different density
ordered SF phases have been proposed for a spin-1 system using SOC,
a specific concern is the possibility to study also the charge
density wave (CDW) Mott insulating phase by employing a spin-1 BHM with non local nearest neighbour extended
interactions apart from the usual onsite interaction, that
may help in realizing the CDW phase. We feel such an extended interaction is relevant in the present context. Although the issues are reasonably well studied in the context of scalar particles \cite{PhysRevA.83.051606,PhysRevA.91.033613,PhysRevB.86.054520,PhysRevLett.94.207202,Apurba}, however it has not been explored for systems with internal degrees of freedom. The
CDW phase which breaks the crystal translational symmetry and thus
have different density modulation corresponding to different
sublattices, forms a new crystalline phase which is also an
incompressible phase like the MI phase defined by an integer occupancy
at each lattice site. The extended interaction, which is
long range in nature may be realized through the dipole-dipole
interaction between the dipolar atoms, not only paves the way for the
CDW phase, but also an additional compressible phase known as the
supersolid (SS) phase which depicts a coexistence of both the crystalline and superfluid
phases.  \\\indent The experimental realization of
$^{52}Cr$ atoms, which have no nuclear spin but have heyperfine spin-3
\cite{PhysRevLett.94.160401,PhysRevLett.106.255303}, has created much interest to study the extended
BHM from a theoretical perspective. Also for small dipole interaction
strengths, the ground state structure of spin-1 dipolar condensate has
been studied through single mode approximation (SMA) in
Ref.\cite{PhysRevLett.93.040403} and different spin textures like
polar core vortices, chiral spin vortex in ferromagnetic case beyond SMA in
Refs.\cite{PhysRevLett.97.020401,PhysRevLett.97.130404}.  \\ \indent
Motivated from such studies on magnetic dipole-dipole interaction which has quite a complicated form, in
this work we study the spin-1 BHM in presence of the density-density interaction term via the mean field approach (MFA). Our plan is to obtain the phase
diagrams for both the AF and the ferromagnetic interactions in presence of extended interaction strengths. We have also performed a perturbation expansion to provide support for the mean field
phase diagrams. We extend our calculations in presence of an external
harmonic confinement and calculate the momentum distribution
corresponding to different phases.  \\\indent This paper is organized
as follows. In section II, we outline our theoretical model for a
spinor Bose gas in presence of an extended interaction described by a BHM and study it via
the familiar mean field technique. In section III, we discuss the
phase diagrams of the system for both MFA and perturbative
approach. Finally, conclusions are drawn in section IV to depict the key results obtained by us.
%%%%%%%%%%%%%%%%%%%%%%%%%%%%%%%%%%%%%%%%%%%
\section{Model}
The BHM for spin-1 ultracold atoms in presence of nearest neighbour
extended interaction can be written as \cite{PhysRevLett.81.742,ohmi,PhysRevA.83.051606},
\begin{eqnarray}
\hat{H}=&-&t\sum\limits_{<i,j>}\sum\limits_{\sigma}(\hat{a}^{\dagger}_{i\sigma}\hat{a}_{j\sigma}+h.c)+\frac{U_{0}}{2}\sum\limits_{i}\hat{n}_{i}(\hat{n}_{i}-1)\nonumber\\ &+&\frac{U_{2}}{2}\sum\limits_{i}({\bf{S}}^{2}_{i}-2\hat{n}_{i})-\sum\limits_{i}\mu_{i}\hat{n}_{i}+V\sum\limits_{<i,j>}\hat{n}_{i}\hat{n}_{j}
\label{bhm1}
\end{eqnarray}
Here $<i,j>$ are the nearest neighbour sites, $t$ is the hopping
amplitude. $\mu_{i}=\mu-V_{ho}$ is the chemical potential at site $i$
and $V_{ho}$ is the trapping potential for harmonic confinement which
has the form $V_{ho}=V_{T}({\bf{x}}-{\bf{x}}_{i})^{2}$ \cite{PhysRevA.76.023606} where
${\bf{x}}=(x,y)$ and ${\bf{x}}_{i}=(x_{i},y_{i})$ are the lattice coordinates at the
center and $i$-site of the trap in a two dimensional square lattice, $V_{T}$ being the strength of such trap. $a^{\dagger}_{i\sigma}(a_{i\sigma})$
is the boson creation (annihilation) operator at a site $i$ and the
particle number operator is $n_{i}=\sum\nolimits_{\sigma}n_{i\sigma}$,
$n_{i\sigma}=a_{i\sigma}^{\dag}a_{i\sigma}$. $U_{0}$ is the spin
independent and $U_{2}$ is the spin dependent on-site interactions
which are related to the scattering lengths, $a_{0,2}$ by $U_{0}=(4\pi
\hbar^{2}/M)((a_{0}+2a_{2})/3)$ and $U_{2}=
(4\pi\hbar^{2}/M)((a_{2}-a_{0})/3)$ corresponding to $S$=0 and $S$=2
channels respectively \cite{PhysRevLett.81.742,ohmi}. The spin
dependent interaction, $U_{2}/U_{0}>0$ is known as the antiferromagnetic
(AF) and $U_{2}/U_{0}\le 0$, is known as the ferromagnetic
interaction. The total spin at a site $i$ is given by,
${\bf{S}}_{i}=a^{\dagger}_{i\sigma}{\bf{F}}_{\sigma\sigma'}a_{i\sigma'}$
where ${\bf{F}}_{\sigma\sigma'}$ are the components of spin-1 matrices
and $\sigma=$+1, 0, -1. The last term includes nearest neighbour
extended interaction with a repulsive strength $V$.  \\ \indent To
decouple both the hopping and the density-density interaction terms,
we use the mean field approximation as given by \cite{PhysRevB.77.014503,kurdestany},
\begin{equation}
\hat{a}^{\dagger}_{i\sigma}\hat{a}_{j\sigma} \simeq \langle
\hat{a}^{\dagger}_{i\sigma} \rangle \hat{a}_{j\sigma}
+\hat{a}^{\dagger}_{i\sigma}\langle \hat{a}_{j\sigma}\rangle-\langle
\hat{a}^{\dagger}_{i\sigma}\rangle \langle \hat{a}_{j\sigma}\rangle
\label{deco1}
\end{equation}
\begin{equation}
\hat{n}_{i\sigma}\hat{n}_{j\sigma} \simeq \langle \hat{n}_{i\sigma}
\rangle \hat{n}_{j\sigma} +\hat{n}_{i\sigma}\langle
\hat{n}_{j\sigma}\rangle-\langle \hat{n}_{i\sigma}\rangle \langle
\hat{n}_{j\sigma}\rangle
\label{deco2}
\end{equation}
where $\langle...\rangle$ denotes the equilibrium value of
an operator. The superfluid order parameter and local density at site
$i$ are defined as,
\begin{equation}
\psi_{i\sigma}= \langle \hat{a}_{i\sigma}\rangle,\hspace{1cm}
\rho_{i\sigma}=\langle \hat{n}_{i\sigma} \rangle
\end{equation}
where $\psi_{i}=\sqrt{\psi^{2}_{i\sigma}}=
\sqrt{\psi^{2}_{i+}+\psi^{2}_{i0}+\psi^{2}_{i-}}$ and
$\rho_{i}=\sum\nolimits_{\sigma}\rho_{i\sigma}$.  \\ Using
Eqs.(\ref{deco1}) and (\ref{deco2}) in the Hamiltonian appearing in Eq.(\ref{bhm1}), its mean field
form can be written as,
\begin{eqnarray}
H^{MF}_{i}&=&\underbrace{-zt\sum\limits_{\sigma}(\phi^{*}_{i\sigma}\hat{a}_{i\sigma}+\phi_{i\sigma}
\hat{a}^{\dagger}_{i\sigma})+zt\sum\limits_{\sigma}\phi^{*}_{i\sigma}\psi_{i\sigma}}_\text{$H^{'}_{i}$}\nonumber\\ 
&+&\underbrace{\frac{U_{0}}{2}\hat{n}_{i}(\hat{n}_{i}-1)+\frac{U_{2}}{2}({\bf{S}}^{2}_{i}-2\hat{n}_{i})}\nonumber\\ &-&\underbrace{\mu_{i}\hat{n}_{i}+zV \bar{\rho_{i}}(\hat{n}_{i}-\rho_{i})}_\text{$H^{0}_{i}$}
\label{mf}
\end{eqnarray}
where $\phi_{i\sigma}=(1/z)\sum\nolimits_{j}\psi_{j\sigma}$ and
$\bar{\rho}_{i\sigma}=(1/z)\sum\nolimits_{j}\rho_{j\sigma}$ and the
sum $j$ includes all the nearest neighbours of the site $i$ of a square
lattice with the coordination number, $z=4$. Here $H^{'}_{i}$ and
$H^{0}_{i}$ are the perturbation term and the unperturbed Hamiltonian
respectively.  \\\indent The presence of an external trapping potential,
$V_{T}$ makes the mean field Hamiltonian $H^{MF}_{i}$ inhomogenous
across the lattice. Thus it necessitates diagonalization of Eq.(\ref{mf}) on a full lattice. Here we have considered a square
lattice of size $L \times L $ and obtain the ground state energy and
eigenfunctions of the system. \\\indent Since the extended interaction term
gives an extra 
\begin{figure*}[!t]
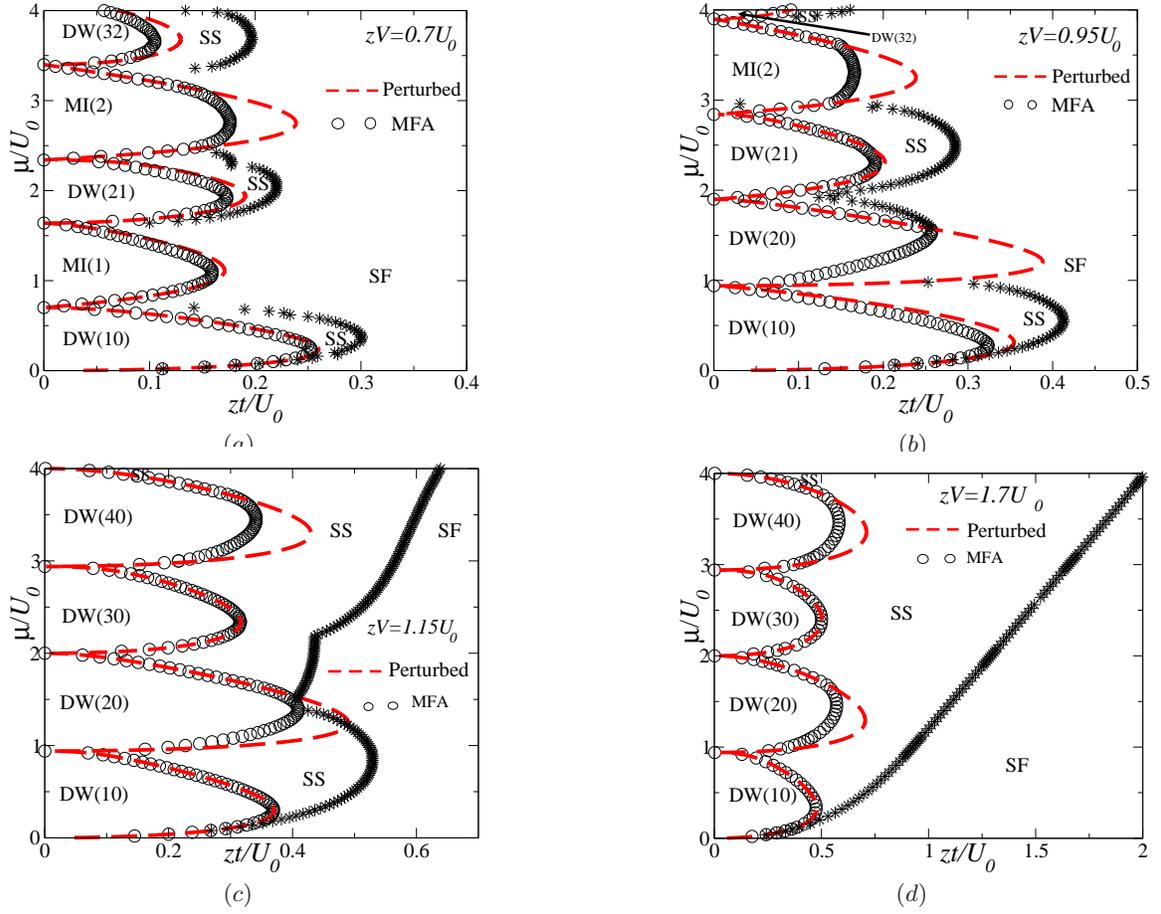

  \centerline{ \hfill
\psfig{file=1a.eps,width=0.35\textwidth}
    \hfill\hfill 
      \psfig{file=1b.eps,width=0.35\textwidth}    
         \hfill} 
\centerline{\hfill $(a)$ \hfill\hfill $(b)$
    \hfill}
\centerline{ \hfill
\psfig{file=1c.eps,width=0.35\textwidth}
    \hfill  \hfill 
      \psfig{file=1d.eps,width=0.35\textwidth}           
         \hfill} 
\centerline{\hfill $(c)$ \hfill\hfill $(d)$ \hfill}
  \caption{Phase diagrams in AF case with $U_{2}/U_{0}=0.03$ for
    different values of $zV/U_{0}$ are shown from (a)-(d). The dotted lines from the perturbation calculation and the circles represent the mean field phase diagrams. At
    $zV/U_{0}=0.7$ [Fig.(a)], phase diagram consists of various CDW
    phases and the even MI(2) and odd MI(1) phases. In Fig.(b), for
    $zV/U_{0}=0.95>1-2U_{2}/U_{0}$, the odd MI(1) lobes now gripped by
    the CDW(20) phase. In Fig.(c), the MI(2) phases now occupied by
    the CDW(40) phase since $zV/U_{0}=1.15>1+2U_{2}/U_{0}$. At strong
    interaction limit, $zV/U_{0}=1.7$, all insulating phase are now
    the CDW phases.}
\label{1}
\end{figure*}
CDW phase which has density modulations as one traverses from one lattice site to another, it is quite reasonable to break the entire lattice into
two sublattices, such as the $A$ and $B$ sublattices. So the unit cell has two types of atoms,
namely $A$ and $B$, where an $A$ atom has $B$ atom as all its neighbours and vice versa. In the CDW phase, one type of sublattice
has higher occupancy than the other, so without much loss of generality, we assume $n_{A}>n_{B}$.
\\\indent To begin with, let us consider $V_{T}=0$. In the absence of trapping, we can diagonalize Eq.(\ref{mf})
over the unit cell consisting of two sites namely $A$ and $B$. Thus for $A$ type of sublattice, Eq.(\ref{mf}) can be written as,
\begin{eqnarray}
H^{MF}_{A}&=&-zt\sum\limits_{\sigma}(\psi_{B\sigma}\hat{a}_{A\sigma}+h.c)+zt\sum\limits_{\sigma}\psi_{B\sigma}\psi_{A\sigma}\nonumber\\&-&\mu_{A}\hat{n}_{A}+\frac{U_{0}}{2}\hat{n}_{A}(\hat{n}_{A}-1)\nonumber\\&+&\frac{U_{2}}{2}({\bf{S}}^{2}_{A}-2\hat{n}_{A})+zV\rho_{B}(\hat{n}_{A}-\rho_{A})
\label{mfA}
\end{eqnarray} 
Similarly one can easily obtain the $H^{MF}_{B}$ by changing the
indices from $A$ to $B$ (and $B$ to $A$) and the total mean field Hamiltonian is just
sum of two mean field Hamiltonians that is
$H^{MF}_{i}=H^{MF}_{A}+H^{MF}_{B}$. The self consistent ground state
energy and the eigenfunctions are obtained by diagonalizing
$H^{MF}_{i}$ in the occupation basis $|\hat{n}_{i\sigma}\rangle $
with $n_{i}=7$ starting with some guess values for $\psi_{(A/B)\sigma}$
and $\rho_{(A/B)\sigma}$ and continue the diagonalization process until
self consistency conditions for the order parameter, $\psi_{i}$ and occupation densities, $\rho_{i}$ are reached.
%%%%%%%%%%%%%%%%%%%%%%%%%%%%%%%%%%%%%%%%%%%%
\section{Results}
\indent It was experimentally found that for $^{23}Na$ atoms, the
spin dependent interaction values are $U_{2}/U_{0}=0.031$
while for $^{87}Rb$ atoms, the same is -0.046 \cite{PhysRevLett.81.742}. Thus here we present our numerical results for different strengths of the
extended interaction, $V$ corresponding to both the AF and ferromagnetic
interactions. The phase diagrams are
calculated based on the self consistent values of the SF order parameters,
$\psi^{eq}_{(A/B)}$ and the local densities, $\rho^{eq}_{(A/B)}$ (we shall drop the superscript, $eq$ from here) to
characterize different phases. \\\indent Both the CDW and MI phases
are incompressible phases with integer occupation densities and
vanishing SF order parameters, however the CDW phase is characterized by unequal occupation densities in the $A$ and $B$ sublattices, that is, $\rho_{A}\ne \rho_{B}$, (the MI phase
corresponds to $\rho_{A}=\rho_{B}$) where $\rho_{A/B}$ is an integer. Also the SF and SS phases are the
compressible phases with non integer densities and finite values of the SF
order parameters but in the SF phase, $\psi_{A}=\psi_{B}\ne 0$ and
$\rho_{A}=\rho_{B} \ne$ integer, while the SS phase is characterized by
$\psi_{A}\ne \psi_{B}\ne 0$ and $\rho_{A}\ne\rho_{B} \ne$ integer
respectively. We present our phase diagrams for four different
values of $zV/U_{0}$ in Fig.\ref{1}(a)-(d). The choice of different $zV/U_{0}$ is justified in the subsequent discussion as we move on and the effect of trapping is included at the end of this section.  
\\ \indent In Fig.\ref{1}(a), the phase diagram corresponding to AF case ($U_{2}/U_{0}=0.03$) presented for $zV/U_{0}=0.7$,
shows that the CDW phase appears in between the MI lobes, and thus a direct
transition from the CDW to the SF is interrupted due to the appearance
of the SS phase. The symbol DW($\rho_{A}\rho_{B}$) implies the CDW
phase having occupation densities, $\rho_{A}$ (corresponding to sublattice
$A$) and $\rho_{B}$ (corresponding to sublattice $B$) with average
occupation density, $\bar{\rho}=(1/2)[\rho_{A}+\rho_{B}]$. The SS phase,
which not only appears along with the DW(21) and DW(32) phases, but also exists
along with the DW(10) phase for $\bar{\rho}<1/2$, as was predicted earlier
through quantum Monte Carlo (QMC) studies in
Ref.\cite{PhysRevB.86.054520}.  The MI(1) and MI(2) are the Mott
insulating phases with occupation densities $\rho_{A}=\rho_{B}=1$ and
$\rho_{A}=\rho_{B}=2$ respectively. The odd and even MI phases which
form the spin nematic and singlet phases are the distinguishing
features of a spinor Bose gas compared to a spin-0 Bose gas and the
stabilization of the singlet phase over the nematic phase has been
studied extensively in Refs.\cite{PhysRevA.70.043628,PhysRevB.77.014503}.  \\\indent In Fig.\ref{1}(b), we found
that corresponding to $zV/U_{0}=0.95$, although the phase diagram consists of all
the compressible and incompressible phases, however, interestingly, the
MI(1) phase is now completely occupied by the DW(20) phase and the
chemical potential widths of the DW phases increase with $zV/U_{0}$,
which is now 0.95 (Fig.\ref{1}(b)) compared to 0.7 in
Fig.\ref{1}(a). \\\indent The disappearance of the MI(1) phase can be understood
by considering the atomic limit, that is, $t=0$ where the system only
consists of the MI and CDW phases. In the atomic limit, the ground state energy $E_{g}(n_{A}n_{B})$ from Eq.(\ref{bhm1}) in the CDW phase is given by,
\begin{eqnarray}
E_{g}(n_{A}n_{B})&=&\frac{U_{0}}{4}\sum\limits_{i=A,B}[n_{i}(n_{i}-1)]-\frac{\mu}{2}\sum\limits_{i=A,B}n_{i}+\frac{U_{2}}{4}\nonumber\\
&&\sum\limits_{i=A,B}[S_{i}(S_{i}+1)-2n_{i}]+\frac{zV}{2}n_{A}n_{B}
\label{atomoic}
\end{eqnarray}
Following the calculations carried out in
Ref.\cite{PhysRevA.83.013605}, we found that the chemical potential
width for the odd MI lobes is $U_{0}-2U_{2}$, while for the even MI
lobes, it is $U_{0}+2U_{2}$. Also we have checked that for
the CDW phase, the chemical potential width is $zV$ using the same assumption
that for the odd occupation densities, we shall consider the spin eigenvalues to be $S=1$ and for even occupation
densities, $S=0$ will be considered. So for $zV/U_{0}=1\pm 2U_{2}/U_{0}$, there are
possibilities of coexistence of different CDW and MI phases because of
the degeneracy in their ground state energies. As a result, at
$U_{2}/U_{0}=0.03$, when $zV/U_{0}>0.94$, the MI(1) phase now gets
absorbed by the DW(20) phase and this applies for the other odd
MI lobes as well for $U_{2}/U_{0}<0.5$.  \\\indent At larger value of the extended interaction, that is for
$zV/U_{0}=1+2U_{2}/U_{0}=1.06$, the MI(2) phase becomes degenerate with
the DW(40) phase and beyond this critical value, all insulating phases
become CDW phases and the SS phase has now significantly expanded with increasing $zt/U_{0}$
[Fig.\ref{1}(c)]. We have also obtained the phase diagrams at stronger
values of the extended interaction strength, that is, $V/U_{0}=1.7$ in Fig.\ref{1}(d), which indicates
that the system is more likely to be in the SS phase compared to the
CDW or the SF phases.
\begin{figure}[!ht]
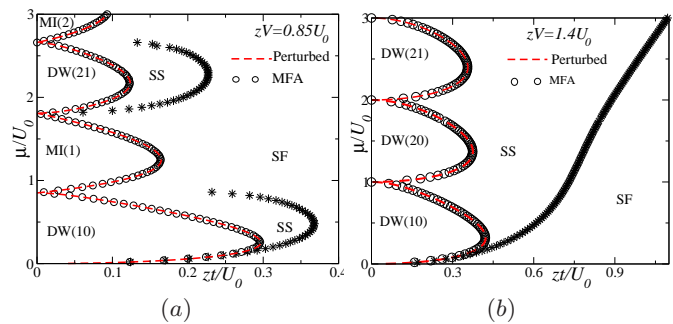

  \centerline{ \hfill
\psfig{file=2a.eps,width=0.25\textwidth}
    \hfill \hspace{-2mm} \hfill 
      \psfig{file=2b.eps,width=0.24\textwidth}           
         \hfill} \centerline{\hfill $(a)$ \hfill\hfill $(b)$
    \hfill}
  \caption{Phase diagrams in the ferromagnetic case with $zV/U_{0}$ for $U_{2}/U_{0}=-0.04$ in (a) and $U_{2}/U_{0}=0.0$ in (b). At $zV/U_{0}=0.85<1+U_{2}/U_{0}$ [Fig.(a)], both the MI and CDW phases exists with the SS and SF phases. In Fig.(b), phase diagram consists only the CDW phases since $zV/U_{0}=1.4>1$.}
\label{2}
\end{figure}
\\\indent In Fig.\ref{2}, we have plotted the phase diagrams for two
different values of $zV/U_{0}$ corresponding to the ferromagnetic
case and found that they show similar
characteristics as that of a scalar Bose gas. In this case, there is no distinction between the
odd and the even MI lobes and hence all the MI lobes have densities,
$\rho_{i}$ with the maximum spin eigenvalue, that is, $S_{i}=\rho_{i}$ \cite{PhysRevA.83.013605}. \\\indent In the
atomic limit, it turns out the the chemical potential width for each of the
MI lobe is $U_{0}+U_{2}$, while for the CDW phase, it is
$zV$. The phase diagram with $zV/U_{0}=0.85$ for $U_{2}/U_{0}=-0.04$ is
shown in Fig.\ref{2}(a) demonstrates all the MI and the CDW
phases alongwith all the compressible phases, since the critical value at
which both the MI and CDW phases become degenerate at
$zV/U_{0}=0.96$. We have also considered a strong interaction limit, namely
$zV/U_{0}=1.4$, which is larger than the critical value 1 for
$U_{2}/U_{0}=0.0$ in Fig.\ref{2}(b). 
\begin{figure*}[!t]
  \centerline{ \hfill
\psfig{file=3a.eps,width=0.22\textwidth}
    \hfill \hfill  
      \psfig{file=3b.eps,width=0.22\textwidth}           
         \hfill
\psfig{file=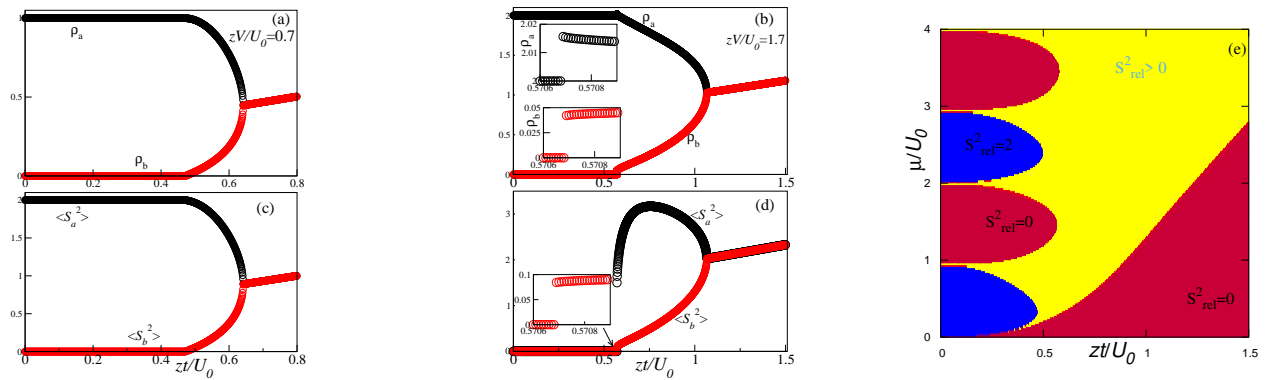,width=0.27\textwidth}} 
  \caption{The 1D behaviour of $\rho_{A/B}$ and $\langle
    S^{2}_{A/B}\rangle$ in the AF case with
    $zV/U_{0}=0.7$ in (a) and $zV/U_{0}=1.7$ in (b). The discontinuity in $\rho_{i}$ and $\langle
    S^{2}_{i}\rangle$ indicate a first order transition for the DW(20)-SS transition (b) and (d)
    while second order for the DW(10)-SS transition respectively in (a) and (c). The
    total spin eigenvalue $\langle S^{2}_{i}\rangle$ is either 2 or 0
    signifies $S_{i}=1$ or $S_{i}=0$ since $\langle
    S^{2}_{i}\rangle=S_{i}(S_{i}+1)$ depending upon the odd or even
    occupation densities corresponding to the respective CDW phase. Relative spin eigenvalues, $S^{2}_{rel}=\langle
S^{2}_{A}\rangle-\langle S^{2}_{B}\rangle$ 
	in AF case for $zV/U_{0}=1.7$ in (e).}
\label{3}
\end{figure*}
Now all the CDW and the
SS phases can be found, however there are no MI lobes. We have also checked for different
values of $zV/U_{0}$ corresponding to $U_{2}/U_{0}=0.0$ and they are in complete
agreement with the results obtained via Gutzwiller approximation in
Ref.\cite{PhysRevA.83.051606}. 
\\\indent Since the formation of spin
singlet pairs corresponding to even occupation densities and their
stabilization over the odd MI lobes has been studied in Refs.\cite{PhysRevA.70.043628,PhysRevB.69.094410,PhysRevB.77.014503} without the
extended interaction, one can ask, is it possible to have also the spin
singlet phase corresponding to integer $\bar{\rho}$ for the CDW
phase. If we look at Fig.\ref{1}(a)-(b), it can be concluded that
although the critical tunneling strength, $zt_{c}/U_{0}$ for transition
from the incompressible to the compressible phases still occurs
at higher values of $zt/U_{0}$ for the MI(2) phase compared to that of the MI(1) phase, but
$zt_{c}/U_{0}$ for the DW(20) phase is now enhanced with increasing
$zV/U_{0}$ compared to the other insulating phases except for the DW(10)
phase. \\\indent Interestingly, from Fig.\ref{1}(c)-(d), we found that the
$zt_{c}/U_{0}$ for a transition from the CDW to the SS phase
corresponding to the DW(20) or DW(40) are higher than that of the
DW(10) or DW(30), thereby indicating a possibility of spin
singlet formation in these CDW phases which we shall confirm by
calculating the local spin eigenvalue in the subsequent discussion.  \\\indent It
is also helpful to study the nature of phase transition for different
phases in the AF case. Earlier studies indicate that
the MI-SF phase transition is first order for the even MI phase and
second order for the odd MI lobes without an extended interaction in
Refs.\cite{PhysRevA.70.043628,PhysRevB.77.014503}. In Fig.\ref{3}(a),
we have shown the one dimensional behaviour of $\rho_{A/B}$ for
$zV/U_{0}=0.7$ at $\mu/U_{0}=0.5$ and it shows a continuous transition
from DW(10) to the SS and then to the SF phases, indicating a second order transition. The same holds true for other DW phases. However the transition from DW to SF phase is found to be first order in nature, and the SF-MI transition for the even and
odd MI phases still show first and second order transition respectively for $zV/U_{0}=0.7$ and $0.95$.  \\\indent We have
checked that for $zV/U_{0}>1.15$, the CDW phases indicate a first or
second order transition depending upon the even or odd occupation densities respectively. There is a discontinuous transition from DW(20) or DW(40) with an average $\bar{\rho}=$ integer to the SS
phase [Fig.\ref{3}(b)], while a continuous transition occurs for the DW(10) or DW(30) with an average $\bar{\rho}\ne$ integer to the SS
phase. The SF order parameters, $\psi_{A/B}$ also
show similar behaviour as that of $\rho_{A/B}$ for different phases
with the extended interaction, $zV/U_{0}$.
\begin{figure*}[!t]
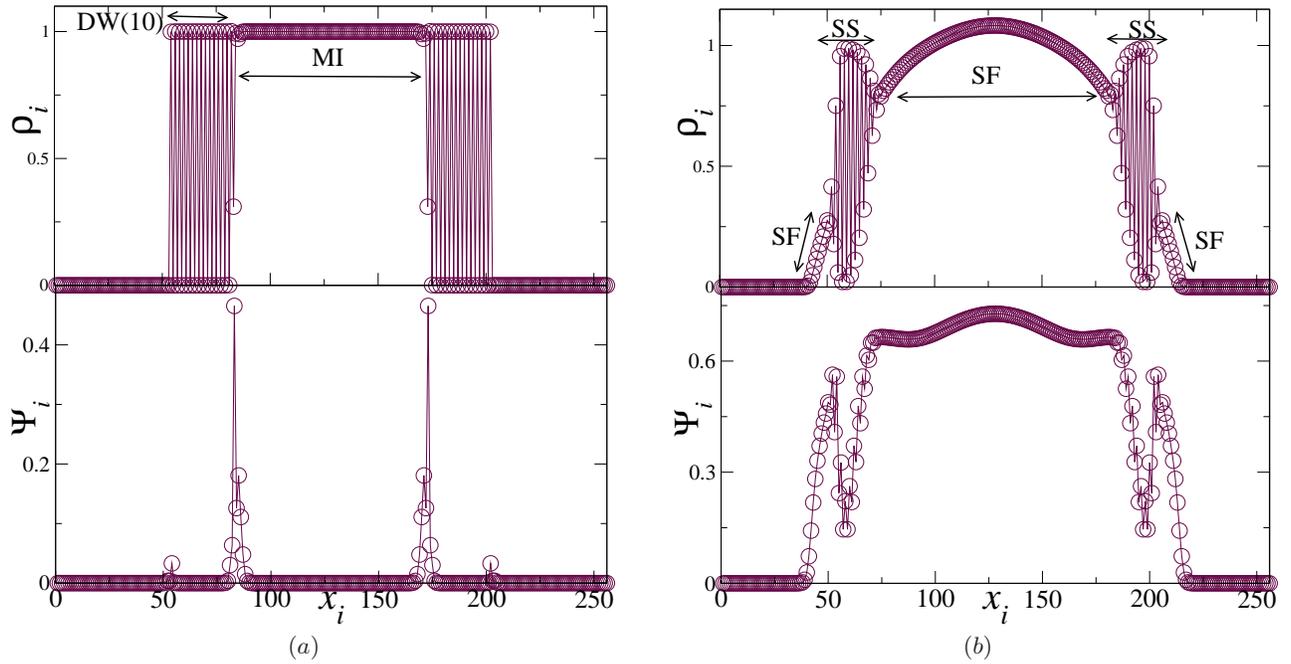

  \centerline{ \hfill
\psfig{file=4a.eps,width=0.45\textwidth}
    \hfill \hspace{-3mm} \hfill 
      \psfig{file=4b.eps,width=0.45\textwidth}           
         \hfill} \centerline{\hfill $(a)$ \hfill\hfill $(b)$
    \hfill}
  \caption{The variation of density, 
$\rho_{i}$ and the order parameter, $\psi_{i}$ in the AF case with $zV/U_{0}=0.7$ and
trapping potential, $V_{T}=0.0002$ for the CDW-MI phase at $\mu/U_{0}=1.1$ and $zt/U_{0}=0.05$ in (a) and 
the SS-SF phase at $\mu/U_{0}=1.25$ and $zt/U_{0}=0.25$ in (b).}
\label{4}
\end{figure*}
\\\indent In Fig.\ref{3}(c) and (d), we have shown the variation of the total spin eigenvalue,
$\langle S^{2}_{i}\rangle$ with different values of $zV/U_{0}$ to verify our claims made in the
previous discussion that the spin eigenvalue to be 0 for the even
occupation densities and 1 for odd occupation densities to calculate
the width of each of the CDW lobes and assess if there is any possibility of spin
singlet formation for the CDW phase having integer $\bar{\rho}$. At
$zV/U_{0}=0.7$, we found that $\langle S^{2}_{A}\rangle=2$ and
$\langle S^{2}_{B}\rangle=0$ for the DW(10) phase [Fig.\ref{3}(c)] and $\langle
S^{2}_{A}\rangle=0$ and $\langle S^{2}_{B}\rangle=2$ for the DW(21)
phase (not shown here). This implies that $S_{A}=1$ for $\rho_{A}=$ odd and $S_{B}=0$
for $\rho_{B}=$ even and vice versa for the CDW phases, since $\langle
S^{2}_{i}\rangle=S_{i}(S_{i}+1)$. We have checked that $\langle S^{2}_{i}\rangle=0$ for the even and $\langle S^{2}_{i}\rangle=2$ for the odd MI phases and the transition to the SF phase still remain first and second order respectively 
\cite{PhysRevA.70.043628,PhysRevB.77.014503}. At $zV/U_{0}=1.7$, both
the $\langle S^{2}_{A}\rangle=\langle S^{2}_{B}\rangle=0$ for DW(20) phase, then followed by a jump to the SS phase [Fig.\ref{3}(d)] and we found an identical behaviour also for the DW(40) phase. While for DW(10) or DW(30) phases, $\langle S^{2}_{A}\rangle=2$, $\langle S^{2}_{B}\rangle=0$ in the CDW phase and show continuous transition to the SS and SF phases. So the DW(20) or DW(40) behave as the spin singlet while the DW(10) or DW(30) as spin nematic CDW insulator phases like the spin singlet and nematic phases corresponding to the even and the odd MI lobes. 
\\\indent
Further we have plotted the relative spin eigenvalue that is $S^{2}_{rel}=\langle
S^{2}_{A}\rangle-\langle S^{2}_{B}\rangle$ with $zt/U_{0}$ for
$zV/U_{0}=1.7$ in Fig.\ref{3}(e). At this value, $S^{2}_{rel}$ shows that, for the
CDW phases, $S_{A/B}$ oscillates between 0 and 1 depending
upon the density variations of these particular phases, that is the blue
and red lobes pertaining to the CDW phase having non integer and
integer $\bar{\rho}$ respectively. The
yellow region is for the SS phase where $S^{2}_{rel}\ne 0$ and then the SF phase corresponds to the red region where $S^{2}_{rel}= 0$. Thus we found signatures of a spin density wave (SDW) pattern, where $S_{i}$ oscillates between 0 and 1 in a similar fashion as that of $\rho_{i}$ in different CDW phases.  \\\indent We shall now turn on our attention to the
second order perturbation calculation to obtain the phase boundaries
between different phases in order to compare them with the mean field
results discussed above. The ground state energy, $E_{n}$ after incorporating the
first order, $E^{(1)}$ and second order, $E^{(2)}$ corrections can be expressed
in terms of $\psi$ and $\phi$ as,
\begin{eqnarray}
\hspace*{-3mm}E_{n}(\psi,\phi)&=&E^{(0)}+E^{(1)}+E^{(2)}\nonumber\\ &\hspace*{-9mm}=&\hspace*{-5mm}E^{(0)}+C_{2}(U_{0},U_{2},\mu,n,V)f(\psi_{\sigma},\phi_{\sigma})
\end{eqnarray}
where $C_{2}$ is the coefficient arising from the perturbation
correction, $f$ includes the order parameter and
$E^{0}$ is the eigenvalue of $H^{0}$ which is given by (site indices are
skipped for the time being),
\begin{eqnarray}
E^{(0)}&=&\frac{U_{0}}{2}n(n-1)-\mu n+\nonumber\\
&&\frac{U_{2}}{2}[S(S+1)-2n]+zV\bar{\rho}(n-\rho)
\end{eqnarray}
Since the order parameter vanishes in the insulating phase, and it remains
finite in the compressible phases, the boundary between them can be
obtained by putting $C_{2}=0$. The phase boundary between the SF-MI phase in the
AF case was obtained earlier by using non degenerate and degenerate
perturbation theory corresponding to the even and odd MI lobes
respectively without $zV/U_{0}$ in
Ref.\cite{PhysRevA.70.043628}. Following similar calculations here, the
above condition for the even MI lobes leads to the following equation,
\begin{equation}
\Big{(}\frac{1}{zt}\Big{)}_{even}=\frac{(n_{i}+3)/3}{\beta_{i}+zV\bar{\rho_{i}}}
-\frac{n_{i}/3}{-\alpha_{i}-2U_{2}+zV\bar{\rho_{i}}}
\label{even}
\end{equation}
while for the odd MI lobes, it is given by,
\begin{eqnarray}
\Big{(}\frac{1}{zt}\Big{)}_{odd}&=&\frac{(n_{i}+2)/3}{\alpha_{i}-zV\bar{\rho_{i}}}+
\frac{4(n_{i}-1)/15}{\alpha_{i}+3U_{2}-zV\bar{\rho_{i}}}\nonumber
\\ &+&\frac{(n_{i}+1)/3}{\beta_{i}-2U_{2}+zV\bar{\rho_{i}}}
+\frac{4(n_{i}+4)/15}{\beta_{i}+zV\bar{\rho_{i}}+U_{2}}
\label{odd}
\end{eqnarray}
where $\alpha_{i}=\mu-(n_{i}-1)U_{0}$ and $\beta_{i}=-\mu+n_{i}U_{0}$
respectively.  Thus the phase boundary between the MI and SF phases in presence of
$zV/U_{0}$ is obtained by using either of Eq.(\ref{even}) or
Eq.(\ref{odd}) separately, depending upon the even or odd MI lobes, with
$\bar{\rho}_{i}=n_{0}$ for $i\in A,B$ since in MI phase,
$\psi_{A}=\psi_{B}=0$ and $\rho_{A}=\rho_{B}=n_{0}$, $n_{0}$ being the
occupancy of that particular MI phase.  \\\indent Now for the CDW-SS
phase boundary, an immediate question arises, namely, which equation one should
use for dealing with the different CDW phases, since it has both even and odd occupation
densities.  As we have seen before, the spin eigenvalue, $S_{i}=0$
corresponding to $\rho_{i}=$ even and $S_{i}=1$ for $\rho_{i}=$ odd in
the CDW phases, we may use a combination of Eq.(\ref{even})
and Eq.(\ref{odd}) depending upon the density to determine the phase boundary with
$\psi_{i}=\psi_{A}$, $\rho_{i}=\rho_{A}$, $\bar{\rho}_{i}=\rho_{B}$ if
$i\in A$ and vice verse for $i\in B$.  For example, for the CDW(10) phase at
$zV/U_{0}=0.7$, $n_{A}=1$, $S_{A}=1$ and $n_{B}=0$, $S_{B}=0$, thus
the boundary equation is given by,
\begin{equation}
\frac{1}{z^{2}t^{2}}=\Big{(}\frac{1}{0.7-x}\Big{)}
\Big{(}\frac{1}{x}+\frac{2}{3(0.97-x)}+\frac{4}{3(1.03-x)}\Big{)}
\label{dw10}
\end{equation}
\begin{figure}[!ht]
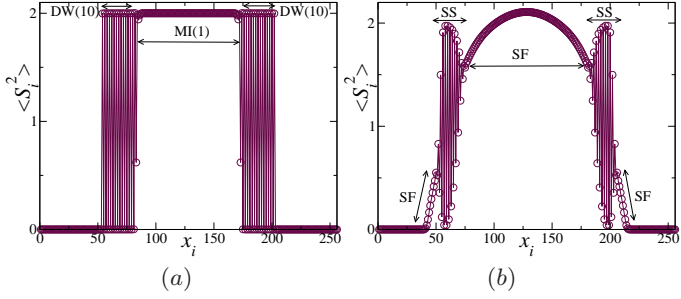

  \centerline{ \hfill 
      \psfig{file=5a.eps,width=0.25\textwidth}           
         \hfill\hspace{-2mm}
	\psfig{file=5b.eps,width=0.25\textwidth}           
         \hfill} 
\centerline{\hfill $(a)$ \hfill\hfill $(b)$
    \hfill}
  \caption{ The variation of $\langle S^{2}_{i}\rangle$ in the AF case with $zV/U_{0}=0.7$ and
trapping potential, $V_{T}=0.0002$ for the CDW-MI phase at $\mu/U_{0}=1.1$ and $zt/U_{0}=0.05$ in (a) and 
the SS-SF phase at $\mu/U_{0}=1.25$ and $zt/U_{0}=0.25$ in (b).}
\label{5}
\end{figure}
\\\indent In a similar fashion, we have calculated the boundary between
all the compressible and incompresible phases using Eq.(\ref{even}) or
Eq.(\ref{odd}) at different values of $zV/U_{0}$ corresponding to the AF case and are
superimposed in Fig.\ref{2} (dotted line). It shows that both the MFA and the
analytic phase diagrams are in accordance with each other for all the CDW
and MI phases well inside the boundary region, small deviation is
observed near the tip of the insulating phases. From
Fig.\ref{1}, we found that the deviation is particularly
prominent for the even occupancies as compared to the odd ones
corresponding to the CDW or the MI phases. This suggests that MFA
fails to tackle the fluctuations properly and holds good only for
small fluctuations, a fact that is very well known.  \\\indent In the ferromagnetic case, we
have done a similar perturbation calculation for the phase boundary for
$U_{2}/U_{0}\le 0$ with maximum spin eigenvalue of $S_{i}$ that is
$S_{i}=n_{i}$. We found that both the MFA and
phase diagrams obtained via perturbed calculations are in complete agreement with each other in 
Fig.\ref{1}. Moreover the resultant boundary equations for the
MI-SF and CDW-SS phase are identical with those obtained for different
values of $zV/U_{0}$ in
Refs.\cite{PhysRevA.79.053634,PhysRevA.83.051606}.
\\\indent So far the results presented above do not include the
trapping potential. Now we shall consider a two dimensional
trapping potential as $V_{ho}=V_{T}[(x-x_{i})^{2}+(y-y_{i})^{2}]$,
where the trap can be chosen at the center of the lattice $L\times L$ that
is at $x=y=L/2$. Since the order parameter and the spin eigenvalues
are now inhomogenous over the lattice sites, here we shall show their
one dimensional behaviour along the $x$-axis as a function of the distance, $x_{i}$ from the center
of the trap by choosing $y=L/2$. Thus we will be able to scan both types of
sublattice simultaneously. \\\indent In Fig.\ref{4}, we have shown density, $\rho_{i}$
and order parameter, $\psi_{i}$ profile in the AF case with a trapping potential,
$V_{T}=0.0002$ for a square lattice of size $L=256$. A careful scrutiny reveals that $V_{T}\sim 10^{-4}$ for a square lattice of dimensions $256 times 256$ will enable us to capture all the different phases and we choose
$\mu/U_{0}$, $zt/U_{0}$ in such a way that the trap center is in the
vicinity of the DW-MI and SS-SF phases respectively. The circles
denote $A$ type and solid lines denote $B$ type sublattice.
\begin{figure}[!ht]
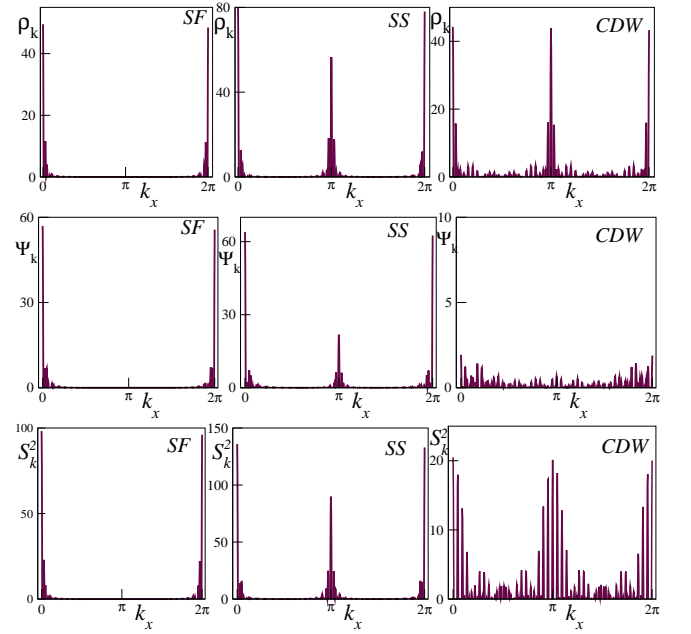

\centerline{ \hfill
\psfig{file=6a.eps,width=0.475\textwidth}
   } 
\centerline{ \hfill   
 \psfig{file=6b.eps,width=0.48\textwidth}    
        \hfill} 
\centerline{ \hfill   
 \psfig{file=6c.eps,width=0.48\textwidth}    
        \hfill} 
  \caption{Magnitude of the Fourier transform of 
$\psi_{i}$, $\rho_{i}$ and $S^{2}_{i}$ with $k_{x}$ in 
SF, SS and CDW phases.}
\label{6}
\end{figure}
\\\indent From Fig.\ref{4}(a), it is clear that they are symmetric about the center of
$x_{i}$ and as one advances from the center in either direction, we find, $\psi_{i}=$ 0
while $\rho_{i}=$ 1 which signals the MI(1) phase. Further movement along the $x$ axis leads to a very narrow
region where $\psi_{i}$ is finite and both $\psi_{i}$, $\rho_{i}$ are oscillatory in
nature implying the presence of the SS phase. As we move with $x_{i}$, this narrow SS phase is now followed by a region where $\psi_{i}$ vanishes but $\rho_{i}$ oscillates between 0 and 1, thereby signifying the presence of the the DW(10) phase and finally lead to a vacuum with vanishing $\psi_{i}$ and $\rho_{i}$ at the edge of the trap.
\\\indent In Fig.\ref{4}(b), we found that around the trap center,
$\psi_{i}$ and $\rho_{i}$ are finite and they are equal for both the sublattices, indicate the signature of the SF phase.
Upon moving away from the SF phase, one encounters the SS phase with oscillating
$\psi_{i}$ and $\rho_{i}$ values which enters into the SF phase and eventually to the vacuum phase. We have also
obtained the density and order parameter dependencies with the lattice site, $x_{i}$ for
different values of $zV/U_{0}$ and $V_{T}$ and they exhibit similar
properties to those discussed above for different phases.  \\\indent The variation of $\langle S^{2}_{i}\rangle$ for different phases in the AF case with the trapping strength, $V_{T}=0.0002$ are shown in Fig.\ref{5}. It shows similar behaviour to that of $\psi_{i}$ and $\rho_{i}$ and found that the MI(1) phase with $\langle S^{2}_{i}\rangle=$ 2 is sandwiched between the DW(10) phases, where $\langle
S^{2}_{i}\rangle$ oscillates between 0 and 2 in Fig.\ref{5}(a). Further an oscillatory behaviour of $\langle
S^{2}_{i}\rangle$ in the SS phase is observed which is in between the SF phases in Fig.\ref{5}(b). \\\indent In the ferromagnetic
case, the variations of $\psi_{i}$ and $\rho_{i}$ are in complete
agreement with the results obtained in Refs.\cite{kurdestany,Apurba1}. We have also checked
the spin eigenvalues and the order parameter profiles in presence of one dimensional trapping potential without $zV/U_{0}$ in both the AF and ferromagnetic cases are in agreement
with the results obtained in Ref.\cite{PhysRevA.76.023606}.  \\\indent Although the order
parameter and the density modulation with lattice sites give an impression
about the different phases but to experimentally realize their signature one has to record the interference pattern via a {\it{time-of-flight}}
experiment. In a {\it{time-of-flight}} experiment, the trapped atoms in
optical lattices are allowed to expand suddenly to register the interference
patterns corresponding to a given Bloch state which is a
superposition of plane waves with a spread in the momentum values. \\\indent In
Fig.\ref{6}, we have shown the Fourier transform amplitude of
$\rho_{i}$, $\psi_{i}$ and $S^{2}_{i}$ with the momentum along the $x$-direction, namely $k_{x}$ in the
SF, SS and CDW phases. The variation of $\rho_{k}$ [Fig.\ref{6}(top)] shows the appearance of peaks at $k_{x}=0$ and $k_{x}=2\pi$ for all the phases, while an additional peak appears at $k_{x}=\pi$ for the CDW or the SS phase. The $\psi_{k}$ [Fig.\ref{6}(middle)] and $S^{2}_{k}$ [Fig.\ref{6}(bottom)] shows similar behaviour as that of the SF and the SS phases except for the CDW phase, no peak is observed in $k_{x}$ while tiny peaks are due to the trapping potential. \\\indent We have also considered the scenario in two dimensions by including $k_{y}$ and found that the peaks in $\rho_{k}$ corresponds to $(k_{x},k_{y})=(2\pi j,2\pi m)$ for the SF or SS
phases in addition to 
$(k_{x},k_{y})=(\pi j,\pi m)$ for the SS or CDW phases where $j,m$ are integers. The momentum
profile of $\psi_{k}$ and $S^{2}_{k}$ shows similar peak positions as that
of $\rho_{k}$ in the SF and SS phases but no peak at $(k_{x},k_{y})=(\pi j,\pi m)$ for CDW phase. In the ferromagnetic case, we have checked that
the Fourier transform profiles are in agreement with those in Refs.\cite{kurdestany,Apurba1}.
\section{Conclusion}
In this work, we have studied the spin-1 BHM in presence of nearest
neighbour extended interactions corresponding to both values of the spin
dependent interactions using the mean field and the perturbation expansion approach. In the AF
case, we have used different justifiable values of the extended interaction
strength corresponding to the odd and
even MI lobes. In the weak interaction limit, the phase diagram
consists of the CDW phase, the MI phase along
with the compressible SF and SS phases. In the strong interaction limit,
when extended interaction is larger than the width of the odd and even MI lobes, all the MI
phases get captured by the CDW phase, since at this critical value,
both the CDW and MI phases become degenerate. Further increase of the
interaction strength leads to the stability of the SS phase over the
other phases.  \\\indent We have also found that the DW-SF phase shows
a first order transition due to a jump in the order parameter. While the DW-SS
phase transition is second order in nature for the odd occupancies and of first order for the even occupation densities. The MI-SF phase transition still remain the first and
second order respectively for the even and odd MI lobes. We have also calculated
the local spin values to confirm the formation of the spin singlet and
nematic CDW phases. In the CDW phase, the spin eigenvalues oscillate between 0 and 1 replicating a spin density wave (SDW)
pattern.  \\\indent Further, we have obtained the phase diagrams through the
perturbation calculation and the boundary between the CDW-SS and MI-SF
phases to compare them with the mean field results. Although the
phase diagrams are in agreement with each other, however small discrepancy
appears at the tip of the insulating phases and particularly it is
prominent for the incompressible phases with even occupation
densities.  \\\indent Also we have studied the order parameter and
spin profile in presence of the trapping potential to characterize
different phases. In order to get a close resemblance with that of the experimental
observations, we have computed the Fourier transform of the order parameter
that demonstrates the appearance of peaks at different momenta values which can
be experimentally observed via {\it{time-of-flight}} experiment. In the
ferromagnetic case, the phase diagrams are similar to the
spin-0 Bose gas and both the mean field and the analytic results are in
excellent agreement with each other.
\\\indent Finally, the extended interaction strength which is closely related with the long range dipole-dipole interaction can be of either electric or magnetic in nature having coupling constant $C_{dd}=\mu^{2}/\epsilon_{0}$ or $C_{dd}=\mu_{0} \mu^{2}$ respectively where $\mu$ being the dipole moment.  For a polarized molecules, the electric dipole moment is very prominent compared to the magnetic dipole moment. But for alkali atoms with spin degrees of freedom, magnetic dipole interaction is notable and very recently for spinor condensates, large magnetic dipole moment of the order of $1\mu_{B}$ for $^{87}Rb$ and $6\mu_{B}$ for $^{52}Cr$ are reported \cite{Lahaye}.
\bibliography{referance} \bibliographystyle{aip}
\end{document}